\documentclass[twocolumn,tightenlines,10pt,amsfonts,amssymb,balancelastpage,
showpacs,superscriptaddress,nofootinbib]{revtex4}%
\usepackage{amsfonts}
\usepackage{amsmath}
\usepackage{amssymb}
\usepackage{subfigure}
\usepackage{graphicx}%
\usepackage{color}

\begin{document}

\title{Quantum coherent control of highly multipartite continuous-variable
entangled states by tailoring parametric interactions}

\author{Giuseppe Patera}\email{Giuseppe.Patera@phlam.univ-lille1.fr}

\affiliation{Laboratoire de Physique des Lasers, Atomes et Mol\'{e}cules,
Universit\'{e}
Lille 1, 59655 Villeneuve d'Ascq Cedex, France, EU}

\author{Carlos Navarrete-Benlloch}

\affiliation{Departament d'\`{O}ptica, Universitat de Val\`{e}ncia, Dr. Moliner
50, 46100
Burjassot, Spain, EU}

\affiliation{Max-Planck-Institut f\"ur Quantenoptik, Hans-Kopfermann-Strasse 1,
85748
Garching, Germany, EU}

\author{Germ\'{a}n J. de Valc\'{a}rcel}

\affiliation{Departament d'\`{O}ptica, Universitat de Val\`{e}ncia, Dr. Moliner
50, 46100
Burjassot, Spain, EU}

\author{Claude Fabre}

\affiliation{Laboratoire Kastler Brossel, Universit\'{e} Pierre et Marie
Curie-Paris6,
ENS, CNRS, 4 place Jussieu CC74, 75252 Paris cedex 05, France, EU}


\begin{abstract}
The generation of continuous-variable multipartite entangled states is
important 
for several protocols of quantum information processing and communication, such
as one-way quantum computation or controlled dense coding. In this article we
theoretically show that multimode optical parametric oscillators can produce a
great variety of such states by an appropriate control of the parametric
interaction, what we accomplish by tailoring either the spatio-temporal shape of
the pump, or the geometry of the nonlinear medium. Specific examples involving
currently available optical parametric oscillators are given, hence showing that
our ideas are within reach of present technology.
\end{abstract}

 % end of PACS codes

\date{\today}
\maketitle

\section{Introduction}

Squeezed states of light were introduced several decades ago as states which
could help beating the limits set by quantum mechanics on the precision of
measurements performed with coherent light \cite{BlueBook,Bachor}. These are states
in which one of the quadratures of light (equivalent to the position and
momentum of a mechanical oscillator) has a quantum uncertainty below the
vacuum or ``shot noise'' level, at the expense of increasing the quantum
fluctuations in the orthogonal quadrature. These states found their way into
the new century thanks to their widely proved applications in ultra-precise
metrology (such as gravitational wave detection \cite{Goda08,Vahlbruch09} or
beam displacements \cite{Treps02,Treps03}), as well as in the field of
quantum information with continuous variables \cite%
{Braunstein05,WeedbrookReview}, where the highest-quality entangled states
(the basic ingredient of many quantum information protocols) known to date are currently
obtained by mixing squeezed beams with linear optics \cite%
{Furusawa98,vanLoock00}.

Squeezed light can be obtained via the parametric down-conversion process
that takes place inside a second order nonlinear medium pumped by a
laser beam \cite{BlueBook}. In order to increase the nonlinear interaction,
it is customary to insert the nonlinear crystal in an optical
cavity---dealing then with a so-called ``optical parametric oscillator''
(OPO)---, and large levels of squeezing are obtained in the down-converted
field when the OPO is operated close to threshold, 93\% of noise reduction
being the current benchmark \cite{Mehmet10} (see also \cite%
{Vahlbruch08,Takeno07}).

On the other hand, quantum information has reached a stage where real-world
applications stimulate an intense research for the implementation of
reliable and practical quantum protocols for quantum communication and
information processing. Several of the promised benefits require though a
quantum ``substrate'' that is created by distributing quantum correlations
(entanglement) among a number of degrees of freedom (modes) increasing with
the complexity of the task to achieve. The protocols of quantum telecloning 
\cite{vanLoock01,Koike06} and controlled dense coding \cite{Zhang02,Jing03}
constitute paradigmatic examples of this scenario for small number of modes,
while one-way quantum computation \cite%
{Raussendorf2001,Menicucci2006,Menicucci08,Gu09}, in which the computation
is achieved by applying local measurements to a set of modes initially in a
cluster state, is a most promising example in the large number of modes
regime.

However, the generation of such multipartite entangled states by means of
optical devices requires experimental configurations whose complexity
increases with the number of modes involved \cite%
{vanLoock00,Jing03,Aoki03,Su07,Yukawa08,Cassemiro08}. In contrast, a
practical source should be compact, scalable, and permit to master
the quantum properties of the generated states even when the number of modes
is very large.

A continuous-variable cluster state source with these properties was
proposed in \cite{Pfister04}---and refined in subsequent papers \cite%
{Bradley05,Menicucci2007} (see also \cite{Menicucci08})---, which consisted
in an OPO driven by a multifrequency pump field in the presence of
concurrent nonlinearities. Recently, these ideas have been brought to the laboratory, and the record of 15 quadripartite cluster states have been generated along 60 consecutive longitudinal modes of the OPO cavity \cite{Physher11}. This interesting approach is however limited by
the number of frequencies that can be coupled within the phase-matching
bandwidth, the need of frequency-sensitive measurements, and the fact that a
nonlinear crystal with different phase-matching properties has to be
engineered for each particular cluster state that one desires to create. An
even more simple and promising approach based on the synchronously operation
of only one single mode vacuum squeezer and a quantum non-demolition gate
was proposed very recently in \cite{Menicucci2010}.

In this paper we propose an alternative approach based on the use of a naturally multimode optical parametric oscillator, either in the spatial \cite
{Lopez2005,Perez06,Perez07,Lopez2008,Navarrete08,ChalopinArXiv,Navarrete2009,Lassen09,Janousek09,Lopez2009,Navarrete10,Chalopin2010}
or in the temporal domain \cite{Valcarcel2006,Patera2010}. Below threshold, they produce multimode squeezed states, which has been observed experimentally \cite{Chalopin2010,Pinel}. When operated
above threshold, such multimode devices have been recently proved to be
capable of producing non-critically squeezed states of light via the
phenomena of spontaneous symmetry breaking \cite%
{Perez06,Perez07,Navarrete08,Navarrete10,Navarrete11} and pump clamping \cite{Navarrete2009,Chalopin2010}.

The advantage of the scheme introduced in the current article for the
generation of cluster states over previous proposals \cite%
{Menicucci08,Pfister04,Bradley05,Menicucci2007} is that one does not need to
engineer the couplings between each mode through crystal design (which is specific of each coupling scheme): choosing instead the temporal or
spatial shape of the pump (which is a flexible technique) is enough to directly control the characteristics of the generated quantum
state, even for a very large number of modes. We will show that it is
possible to master in this way the number of entangled modes, as well as the
distribution of their quantum correlations, thus permitting the generation
of arbitrary multimode Gaussian quantum states of any dimension. The technique that we propose is therefore reminiscent of the widely used ``coherent control" of evolution of atoms and molecules that is obtained by appropriately shaping the pulses with which they interact \cite{Suchowski08,ShapiroBook}.

The article is organized as follows. In Section \ref{Section2} we show how
both the light generated via single-pass spontaneous parametric down
conversion in the low-gain regime, as well as that leaving a multimode
optical parametric oscillator below threshold, can be described as a combination of
independent, squeezed modes (termed ``supermodes'' later on). After
considering a particular (but physically relevant) example of the parametric
interaction to show how these supermodes arise (Section \ref{Section4}), we
demonstrate that the squeezing and spatio-temporal shape of such modes can be
controlled either by tailoring the shape of the pumping field, or the
geometry of the nonlinear medium; we consider separately the cases of an OPO
with either many longitudinal modes (Section \ref{ContinuousCase}) or just a
few transverse modes (Section \ref{DiscreteCase}) available for parametric
down-conversion. We then explain how our results can be used for the
generation of arbitrary cluster states of copropagating modes in Section \ref%
{Cluster}, and give our conclusions in Section \ref{Conclusions}.

\section{Generalized supermode description of weak parametric interactions}

\label{Section2}

The dynamics of parametric interactions in the low-gain regime (for
single-pass devices) or below its oscillation threshold (for cavity devices) is controlled by a kernel function or matrix (depending on the
continuum or discrete nature of the modes involved) which describes the
coupling among the different relevant signal-idler modes. As shown in
previous studies, the diagonalization of such kernel is instrumental for
determining the objects with well-defined quantum properties, which turn out
to be linear combinations of the signal-idler modes. Such linear
combinations were termed ``supermodes" in \cite{Valcarcel2006} and the next
subsections are devoted to their introduction in both single-pass and cavity
devices. We will consider collinear type I degenerate phase matching for
definiteness but our treatment can be generalized easily to other types of
phase matching.

\subsection{Single-pass parametric interactions}

In the single-pass configuration the pump beam amplifies parametrically the
quantum noise impinging the crystal around the subharmonic frequencies and
the down-converted field is measured at the exit of that crystal. In the
undepleted pump approximation, holding when parametric gain is low, the
``output" boson operators are generically related to the ``input" ones via
the well known Bogoliubov transformation 
\begin{equation}
\hat{a}_{\mathrm{out}}(\xi )=\int \mathrm{d}\xi ^{\prime }\left[ C\left( \xi
,\xi ^{\prime }\right) \hat{a}_{\mathrm{in}}\left( \xi ^{\prime }\right)
+S\left( \xi ,\xi ^{\prime }\right) \hat{a}_{\mathrm{in}}^{\dag }\left( \xi
^{\prime }\right) \right] ,  \label{bogoliubov_gen}
\end{equation}%
where $\xi $ is a general continuous index standing for frequency or/and
transverse position\footnote{%
Transverse means orthogonal with respect to a propagation direction, in
which case the paraxial approximation is assumed.}, and $C$ and $S$ are
Green functions that solve the propagation equations (see \cite%
{Wasilewski2006} for details). In the very weak conversion limit a
perturbative treatment can be applied to first order in the crystal length,
which physically means considering only generation of single photon pairs.
In such case $C\left( \xi ,\xi ^{\prime }\right) =\delta \left( \xi -\xi
^{\prime }\right) $ and $S\left( \xi ,\xi ^{\prime }\right) =gK\left( \xi
,\xi ^{\prime }\right) $, where $g\,$\ is a coupling constant proportional
to the length $l_{\mathrm{c}}$ and nonlinear susceptibility $\chi ^{\left(
2\right) }$ of the crystal and to the square root of the total pump
irradiance $P$, and\textit{\ }$K\left( \xi ,\xi ^{\prime }\right) $ is a 
\textit{kernel}. Hence in this regime (\ref{bogoliubov_gen}) becomes 
\begin{equation}
\hat{a}_{\mathrm{out}}(\xi )\approx \hat{a}_{\mathrm{in}}(\xi )+g\int 
\mathrm{d}\xi ^{\prime }K(\xi ,\xi ^{\prime })\hat{a}_{\mathrm{in}}^{\dag
}\left( \xi ^{\prime }\right).  \label{bogoliubov}
\end{equation}%
As we will see immediately, the form of $K$ is universal, and is given by the
product of the pump amplitude and some function describing the overlap
between modes over the crystal. However, the special form of $K$ depends on
the considered problem and in the following we treat successively the
temporal and spatial aspects of single-pass parametric interaction in the
single photon pair approximation (\ref{bogoliubov}).

Let us begin by the spectral/temporal aspects in which case $\xi =\omega $
corresponds to the frequency of the monochromatic modes (we call this the
temporal case). Assuming that all the parametrically coupled modes have the
same transverse dependence, and neglecting diffraction inside the crystal,
the coupling kernel $K(\omega ,\omega ^{\prime })$ turns out to be \cite%
{Wasilewski2006}%
\begin{equation}
K(\omega ,\omega ^{\prime })=\alpha _{\mathrm{p}}\left( \omega +\omega
^{\prime }\right) D\left( \omega ,\omega ^{\prime }\right) ,
\label{freq coupling}
\end{equation}%
$\alpha _{\mathrm{p}}\left( \omega \right) $ being proportional to the
spectral pump amplitude at frequency $\omega $, and $D$ the usual
phase-matching function\footnote{%
In \cite{Wasilewski2006} a factor $\exp \left\{ \mathrm{i}\left[ k\left(
\omega \right) -k\left( \omega ^{\prime }\right) \right] l_{\mathrm{c}%
}/2\right\} $ is included in the expression (\ref{sinc}). Here it is absent
because we are implicitly working in the interaction picture with respect to
time and space, analogous to Eq. ($17$) in \cite{Wasilewski2006}, see \cite%
{Valcarcel2006,Patera2010} for details.} 
\begin{equation}
D(\omega ,\omega ^{\prime })=\frac{1}{l_{\mathrm{c}}}\int_{-l_{\mathrm{c}%
}/2}^{l_{\mathrm{c}}/2}\mathrm{d}z\,\mathrm{e}^{\mathrm{i}\Delta k\left(
\omega ,\omega ^{\prime }\right) z}=\mathrm{sinc}\left[ \Phi \left( \omega
,\omega ^{\prime }\right) \right] ,  \label{sinc}
\end{equation}%
where $\Delta k\left( \omega ,\omega ^{\prime }\right) =k\left( \omega
+\omega ^{\prime }\right) -k\left( \omega \right) -k\left( \omega ^{\prime
}\right) $, $k\left( \omega \right) $ is the optical wavenumber at frequency 
$\omega $ inside the crystal, $\mathrm{sinc}\left( x\right) =\sin \left(
x\right) /x$ is the sinus cardinal function, and%
\begin{equation}
\Phi \left( \omega ,\omega ^{\prime }\right) =\Delta k\left( \omega ,\omega
^{\prime }\right) l_{\mathrm{c}}/2,  \label{Phi}
\end{equation}%
is a phase mismatch.

Let us now turn to the spatial case within the paraxial approximation, in
which case $\xi =\mathbf{r}$ is the transverse spatial variable ($\mathrm{d}%
\xi =\mathrm{d}^{2}\mathbf{r}$). We define this \textquotedblleft spatial
case" in the sense that pump is assumed monochromatic and focus is put on
the signal/idler (multimode) field at just the subharmonic frequency, in
which case the kernel is given by \cite{Lopez2009}: 
\begin{equation}
K(\mathbf{r},\mathbf{r}^{\prime })=\alpha _{\mathrm{p}}\left( \frac{\mathbf{r%
}+\mathbf{r}^{\prime }}{2}\right) \Delta \left( \mathbf{r}-\mathbf{r}%
^{\prime }\right) ,  \label{space coupling}
\end{equation}%
$\alpha _{\mathrm{p}}\left( \mathbf{r}\right) $ being the normalized pump
amplitude at transverse point $\mathbf{r}$, and $\Delta (\mathbf{r})$ the
diffraction function 
\begin{eqnarray}
\Delta (\mathbf{r}) &=&\frac{\mathrm{i}k_{\mathrm{s}}}{4\pi l_{\mathrm{c}}}%
\int_{-l_{\mathrm{c}}/2}^{+l_{\mathrm{c}}/2}\frac{\mathrm{d}z}{z}\mathrm{%
\exp }\left( \frac{\mathrm{i}k_{\mathrm{s}}}{4z}|\mathbf{r}|^{2}\right) 
\label{Delta} \\
&=&\frac{1}{\pi l_{\mathrm{coh}}^{2}}\left[ \frac{\pi }{2}-\mathrm{Si}\left(
\left\vert \frac{\mathbf{r}}{l_{\mathrm{coh}}}\right\vert ^{2}\right) \right]
,  \notag
\end{eqnarray}%
$k_{\mathrm{s}}$ being the phase-matched signal wavenumber inside the
crystal, $l_{\mathrm{coh}}=\sqrt{2l_{\mathrm{c}}/k_{\mathrm{s}}}$ the
\textquotedblleft coherence length", and $\mathrm{Si}\left( z\right)
=\int_{0}^{z}\mathrm{sinc}\left( u\right) \mathrm{d}u$ the sine integral
function.

In the general case, the spatial and spectral aspects of the
parametric interaction are simultaneously present, giving rise to new
interesting features \cite{Gatti2009} which we will not consider here.

\subsection{Intracavity parametric interactions}
\label{Intracavity}

When an optical cavity is used to enhance the efficiency of the nonlinear
process, the previous nonlinear couplings represented by kernel $K$ are
projected onto the cavity modes, and the multi-dimensional spectral and
spatial properties of the correlated photons can be lost due to such
filtering. However for a given length and geometry of the cavity, which we
assume to have a cylindrical symmetry around the optical axis, a great
number of modes can be simultaneously sustained: in the
frequency domain they are the series of longitudinal modes separated by the
free spectral range of the cavity; in the transverse spatial domain they are
the set of Laguerre-Gauss modes $\{\mathrm{TEM}_{pl}\}_{l \in \mathbb{Z}}^{p \in \mathbb{N}}$.

In the temporal case we assume the pump consisting of an unlimited series of
pulses at a given repetition rate, equal to the free spectral range of the
cavity: this is what is called a Synchronously Pumped OPO (SPOPO) \cite{Valcarcel2006,Patera2010}.
The pump spectrum is thus a frequency comb, consisting of a large number of
frequency components, each of which gives rise to signal-idler photons
belonging to different longitudinal modes through the parametric down
conversion process, and selected by energy and linear momentum conservation.

In the spatial case, on the contrary, the pump is assumed monochromatic with
a given spatial profile. In this case, owing to the linearity of the
interaction in the below threshold regime, one can focus on, say, the
subharmonic signal photons. However these photons belong, in general, to
different transverse modes $\mathrm{TEM}_{pl}$ (degenerate in frequency),
characterized by a constant value of the sum $f=2p+|l|$, known as ``family
index" \cite{Navarrete2009}. As well, even several families of transverse
modes can become relevant in the case of a degenerate cavity such as the the
confocal \cite{Lopez2005,Lopez2008,Garcia2009} or the self-imaging cavity 
\cite{Lopez2009,ChalopinSHG}, in which many different families (with different
family indices $f$) resonate at the same frequency. We will use a single
generic index to label this discrete series of modes.

It is instructive in this intracavity interaction problem to write down the
interaction Hamiltonian in the undepleted pump approximation, which can be
written in general as 
\begin{equation}
\hat{H}_{\mathrm{I}}=\mathrm{i}\frac{\hbar g}{2}\int \mathrm{d}\xi \mathrm{d}%
\xi ^{\prime }K\left( \xi ,\xi ^{\prime }\right) \hat{a}^{\dagger }\left(
\xi \right) \hat{a}^{\dagger }\left( \xi ^{\prime }\right) +\mathrm{H.c.,}
\label{GenHam}
\end{equation}%
where the continuous boson operators $\hat{a}(\xi )$ satisfy the standard
commutation relations
\begin{subequations}
\begin{eqnarray}
\lbrack \hat{a}(\xi ),\hat{a}(\xi ^{\prime })]&=&[\hat{a}^{\dagger }(\xi ),%
\hat{a}^{\dagger }(\xi ^{\prime })]=0,
\\
\lbrack \hat{a}(\xi ),\hat{a}^{\dagger }(\xi ^{\prime })]&=&\delta (\xi -\xi
^{\prime }).
\end{eqnarray}
\end{subequations}
and the kernel $K\left( \xi ,\xi ^{\prime }\right) $ plays a role analogous
to that in single-pass devices.

In the temporal/spectral case (SPOPO) the Hamiltonian of the nonlinear
interaction can be written in terms of discrete creation and annihilation
operators of the cavity longitudinal modes: 
\begin{equation}
\hat{H}_{\mathrm{I}}=\mathrm{i}\frac{\hbar g}{2}\sum_{i,j}K_{ij}\hat{a}%
_{i}^{\dagger }\hat{a}_{j}^{\dagger }+\mathrm{H.c.}  \label{eq:Hsinglemode}
\end{equation}%
where $K_{ij}$ are the specialization of the continuous-variable kernel $K$ (%
\ref{freq coupling}--\ref{Phi}) to the relevant longitudinal modes around
the subharmonic, and $\hat{a}_{i}$ is the boson operator for the
signal-idler mode $i$ verifying $[\hat{a}_{i},\hat{a}_{j}^\dagger] =\delta _{i,j}$. The dynamics of the intracavity modes is then given
by the following quantum Langevin equations 
\begin{equation}
\frac{\mathrm{d}\hat{a}_{i}}{\mathrm{d}t}=-\gamma \hat{a}_{i}+\sqrt{2\gamma }%
\hat{a}_{i,\mathrm{in}}+\gamma \sigma \sum_{j}K_{ij}\hat{a}_{j}^{\dagger }
\label{eq:Langevinmultimode}
\end{equation}%
where $\gamma $ is the cavity loss rate (assumed identical for all
signal/idler modes -- which, by the way, cannot be distinguished from one
another in this collinear type I degenerate phase matching case), and $%
\sigma ^{2}=\left( g/\gamma \right) ^{2}$ is a dimensionless pumping
parameter proportional to the actual pump power (see, e.g. \cite{Patera2010}
for more details).

In the spatial case (as defined in the previous Section), under
circumstances in which only a few signal modes are relevant \cite%
{Navarrete2009}, Hamiltonian (\ref{eq:Hsinglemode}) holds by identifying the
indices $(i,j)$ with the available TEM$_{pl}$ modes at the signal frequency,
whose dynamics are ruled by the quantum Langevin equations (\ref%
{eq:Langevinmultimode}). In degenerate cavities, such as confocal or
self-imaging cavities, a discrete representation is still possible, but a
continuous one (\ref{GenHam}) is more helpful, in which case the quantum
Langevin equations become \cite{Lopez2009}%
\begin{equation}
\frac{\partial \hat{a}\left( \mathbf{r}\right) }{\partial t}=-\gamma \hat{a}%
\left( \mathbf{r}\right) +\sqrt{2\gamma }\hat{a}_{\mathrm{in}}\left( \mathbf{%
r}\right) +\gamma \sigma \int \mathrm{d}^{2}\mathbf{r}^{\prime }K\left( 
\mathbf{r},\mathbf{r}^{\prime }\right) \hat{a}^{\dag }\left( \mathbf{r}%
^{\prime }\right).
\end{equation}%
For self-imaging cavities \cite{Lopez2009} the kernel is given by (\ref{space coupling}) and (\ref{Delta}), while for the confocal case it has a slightly different expression \cite{Lopez2005} owed to the fact that even and odd transverse families resonate at different frequencies.

\subsection{Supermodes}

We have seen in the previous Section that the output from parametric devices
in the low-gain (or below threshold) regime is governed by a kernel $K$,
which is equal, under equivalent conditions, in the single-pass and
intracavity cases. As both the input-output relations in single-pass
configurations and quantum Langevin equations in cavity devices are linear
in the considered regime, diagonalization of the kernel allows a
considerable simplification in the description of the problem as well as a
clear physical picture of the entities in which clean quantum properties are
concentrated. This approach \cite%
{Lopez2009,Valcarcel2006,Patera2010,Wasilewski2006} is the single-partite
and continuous-variable version \cite{Bennink2002} of the Schmidt
decomposition for the bi-partite bi-photon wave function \cite{Law2000},
which in general is formally expressed by the Bloch-Messiah decomposition 
\cite{Braunstein2005}.

Let us introduce the eigenmodes of the continuous kernel $K(\xi ,\xi
^{\prime })$, which are the solutions $s_{n}(\xi )$ of the Fredholm integral
equation: 
\begin{equation}
F\left[ s_{n}\left( \xi \right) \right] \equiv \int \mathrm{d}\xi ^{\prime
}\,K(\xi ,\xi ^{\prime })s_{n}(\xi ^{\prime })=\Lambda _{n}\,s_{n}(\xi ).
\label{Fredholm}
\end{equation}

Starting from the eigenmodes $s_{n}$, we can define the associated \textit{%
supermode} annihilation operator $\hat{S}_{n}$ as 
\begin{equation}
\hat{S}_{n}=\int \mathrm{d}\xi \,s_{n}(\xi )\,\hat{a}(\xi ),
\label{supermode}
\end{equation}
where we assume $s_{n}(\xi )$ to be normalized so that $[ \hat{S}_{n},\hat{S}_{m}^{\dag }] =\delta _{n,m}$. These operators annihilate a
photon in a mode which is a combination of either spatial or frequency
modes, and we call it ``supermode'' \cite{Valcarcel2006} because it is the
combination of many cavity modes and has a non trivial spatial or
spectral/temporal shape in general. Using (\ref{supermode}) and (\ref{Fredholm}) into
Eq. (\ref{bogoliubov}) we get%
\begin{equation}
\hat{S}_{\mathrm{out},n}=\hat{S}_{\mathrm{in},n}+g\Lambda _{n}\hat{S}_{%
\mathrm{in},n}^{\dag }.
\end{equation}

When the crystal is inserted in an optical cavity---OPO case---, Ref. \cite%
{Valcarcel2006} shows that the supermodes can be defined as a discrete
combination of the cavity eigenmodes instead of the continuous combination
written above, corresponding to the diagonalization of matrix $K$ in Eq. (%
\ref{eq:Langevinmultimode}). These two possible expressions are identical in
the limit where the cavity modes are tight enough so that the sum can be
assimilated to an integral. In the following we assume that this condition
is fulfilled. In this case, the evolution of these supermode operators is
governed by a set of uncoupled quantum Langevin equations obtained by
introducing the discrete versions of (\ref{supermode}) and (\ref{Fredholm})
in Eq. \eqref{eq:Langevinmultimode}: 
\begin{equation}
\frac{\mathrm{d}\hat{S}_{n}}{\mathrm{d}t}=-\gamma \hat{S}_{n}+\sqrt{2\gamma }%
\,\hat{S}_{\mathrm{in},n}+\gamma \sigma \Lambda _{n}\hat{S}_{n}^{\dag }.
\label{dSdt}
\end{equation}%
This equation shows that according to this definition, any multimode OPO can
be in fact seen as a set of independent single-mode OPOs, and therefore can
produce a set of copropagating squeezed supermodes. Hence each supermode is
independently squeezed, and the squeezed quadrature can be shown to have a noise level
at zero frequency equal to $(\left\vert \Lambda
_{1}\right\vert -\left\vert \Lambda _{n}\right\vert )^{2}/(\left\vert
\Lambda _{1}\right\vert +\left\vert \Lambda _{n}\right\vert )^{2}$  ($1$ setting the shot-noise level) when the
system is operated close to threshold, where $\Lambda _{1}$ is the eigenvalue
of largest absolute value.

\section{A general method for tailoring the supermode spectrum}

\label{Section4}

The solution of Eq. \eqref{Fredholm} is simple when the kernel factorizes, 
\begin{equation}
K(\xi ,\xi ^{\prime })=f(\xi )f(\xi ^{\prime }),  \label{factor}
\end{equation}%
in which case the unique eigenmode is, within a multiplicative factor, $%
s_{1}=f$, with corresponding eigenvalue $\Lambda _{1}=\int \mathrm{d}\xi
^{\prime }\,f\left( \xi ^{\prime }\right) ^{2}$. This result can be extended
to the case where there is a set of $N$ orthogonal functions $\left\{
f_{n}\right\} _{n=1}^{N}$ such that 
\begin{equation}
K(\xi ,\xi ^{\prime })=\sum_{n=1}^{N}f_{n}(\xi )f_{n}(\xi ^{\prime }),
\label{factorize}
\end{equation}%
in which case Eq. \eqref{Fredholm} has $N$ solutions $s_{n}=f_{n}$ (save
multiplicative constant), $\Lambda _{n}=\int \mathrm{d}\xi ^{\prime
}f_{n}\left( \xi ^{\prime }\right) ^{2}$. When $K$ has the same type of
decomposition as \eqref{factorize}, but with non orthogonal functions, just
linearly independent, one can show that \eqref{Fredholm} has still $N$
solutions \cite{Courant}, which are now different from the functions $f_{n}$.

The analytical description of the kernel $K$ in terms of a basis of linearly
independent functions (supermodes) like in Eq. \eqref{factorize} has a very
simple solution and physical interpretation in the case where the kernel can
be factorized as 
\begin{equation}
K(x,x^{\prime })=K_{+}(x+x^{\prime })K_{-}(x-x^{\prime }).  \label{FG}
\end{equation}%
In the spatial case this is the form encountered \cite{Lopez2009}, see (\ref%
{space coupling}): $K_{+}$ relates to the pump, while $K_{-}$ has to do with
the crystal. In the temporal case, while this is not the most general kernel
one can find in actual applications, it has been shown in \cite%
{Patera2010,Wasilewski2006} that it is a sensible approximation to many real
cases. The function  $K_{+}$ in (\ref{FG}) can be
manipulated by tailoring the pump shape, while in order to tune the function $K_{-} $ one needs to play with the geometry of the nonlinear crystal.

We now exhibit a series of ``simple" kernels which allow a high degree of
control over their spectra and, as we will show in the next Section, find
applications in actual systems.

Let us start by the simplest case, the symmetric Gaussian kernel, defined by 
$K_{+}(x)=K_{-}(x)=$ $\mathrm{e}^{-\frac{1}{2}\sigma ^{2}x^{2}}$, reading 
\begin{equation}
K(x,x^{\prime })=\mathrm{e}^{-\sigma ^{2}x^{2}}\mathrm{e}^{-\sigma
^{2}x^{\prime 2}},  \label{single mode}
\end{equation}%
as in (\ref{factor}), whose only supermode is the Gaussian function $%
s_{1}\left( x\right) =\mathrm{e}^{-\sigma ^{2}x^{2}}$, with corresponding
eigenvalue $\Lambda _{1}=\sqrt{\pi /2\sigma ^{2}}$.

\begin{figure*}[t]
\centering
\includegraphics[width=0.88\textwidth]{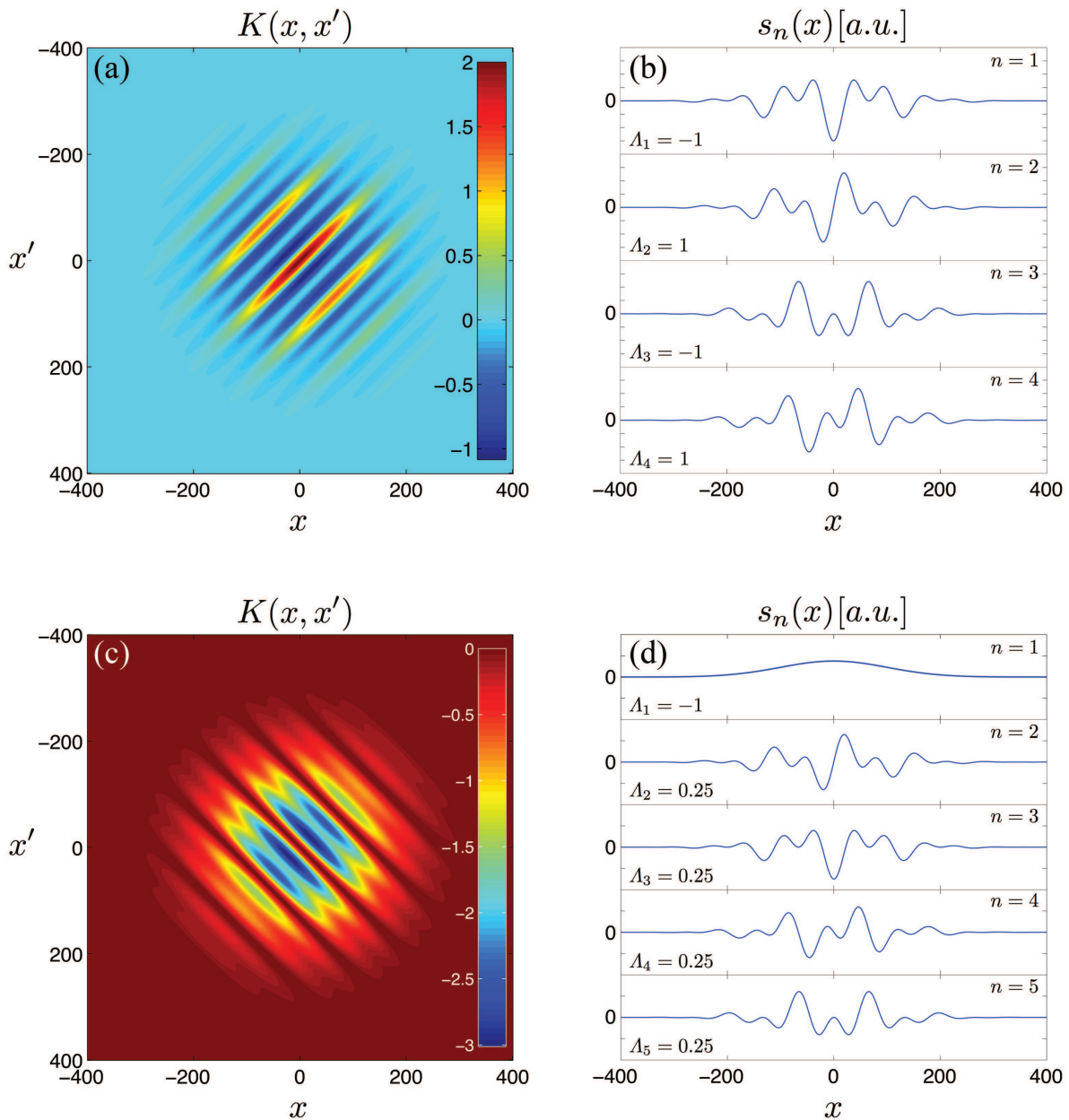}
\caption{(a) Density plot of the kernel---see (\ref{FG}) and (\ref{Gaussian
modulated})---particularized to
the following parameters: $\protect\sigma=0.005$, $b_1^+=1$,
$\protect\beta_1^+=3\protect\pi\protect\sigma$, $b_2^+=1$, $b_0^-=1$
$\protect\beta_2^+=6\protect\pi\protect\sigma$, the rest of $b$'s being zero;
in (b) we show its associated eigenvectors with the corresponding eigenvalues.
(c) Density plot of the kernel for a different choice of the parameters:
$\protect\sigma=0.005$, $b_0^-=-2$, $b_1^-=1$, $b_0^+=1$,
$\protect\beta_1^-=3\protect\pi\protect\sigma$, $b_2^-=1$,
$\protect\beta_2^-=6\protect\pi\protect\sigma$, the rest of $b$'s being zero;
in (d) we show the corresponding eigenvectors and eigenvalues.}
\label{cases12}
\end{figure*}

Let us now consider the previous symmetric kernel (\ref{single mode}), now
multiplied by cosine functions of $x+x^{\prime }$ and $x-x^{\prime }$, i.e.:
\begin{equation}
K_{\pm }\left( x\right) =\mathrm{e}^{-\frac{1}{2}\sigma
^{2}x^{2}}\sum_{n=0}^{N_{\pm }}\,b_{n}^{\pm }\cos \left( \beta _{n}^{\pm
}x\right) ,  \label{Gaussian modulated}
\end{equation}%
which is a quasiperiodic modulation of the Gaussian kernel. In (\ref%
{Gaussian modulated}), $\left\{ b_{n}^{\pm },\beta _{n}^{\pm }\right\}
_{n=0}^{N}$ are constants and $\beta _{0}^{\pm }=0$ by definition. This
leads to a kernel $K$ (\ref{FG}) which can be written like in Eq. %
\eqref{factorize}---with non-orthogonal functions $f_n$---using well known trigonometric formulae. We have therefore
shown that a kernel having the form (\ref{FG}) with (\ref{Gaussian modulated}%
) will have exactly $(2N_{+}+1)(2N_{-}+1)$ eigenmodes, or $4N_{+}N_{-}$
modes if the constant terms $b_{0}^{\pm }=0$. As shown in Appendix A a
simple way to diagonalize the kernel defined by (\ref{Gaussian modulated})
consists in using auxiliary functions%
\begin{equation}
z_{n_{1},n_{2}}^{t_{1},t_{2}}(x)=\mathrm{e}^{-\sigma ^{2}x^{2}}t_{1}\left(
\beta _{n_{1}}^{+}x\right) t_{2}\left( \beta _{n_{2}}^{-}x\right) ,
\label{aux}
\end{equation}%
where $t_{i=1,2}$ stands for any of the trigonometric functions $\cos $ or $%
\sin $, in terms of which actual eigenvectors and eigenvalues can be found
by linear combinations of them. The expressions for eigenvalues and
eigenvectors are awfully cumbersome but one can show that whenever 
\begin{equation}
\left( \beta _{n_{1}}^{+}\right) ^{2},\left( \beta _{n_{2}}^{-}\right)
^{2}\gg 8\sigma ^{2},  \label{approx}
\end{equation}%
the functions (\ref{aux}) are approximate eigenvectors indeed. For each
couple $\left( n_{1},n_{2}\right) $ in (\ref{aux}) two doubly degenerate
eigenvalues are found, of equal magnitude but opposite sign: $\pm \sqrt{\pi
/32\sigma ^{2}}b_{n_{1}}^{+}b_{n_{2}}^{-}$. The positive one is associated
to $t_{1}=\cos $ and the negative one to $t_{1}=\sin $. As for the cases $%
\left( n_{1}=0,n_{2}\right) $ and $\left( n_{1},n_{2}=0\right) $, in which $%
\beta _{n_{1}}^{+}=0$ and $\beta _{n_{2}}^{-}=0$ respectively, the
eigenvalues read $\sqrt{\pi /8\sigma ^{2}}b_{0}^{+}b_{n_{2}}^{-}$ and $-\sqrt{\pi /8\sigma ^{2}}b_{0}^{+}b_{n_{2}}^{-}$, respectively. Finally, the eigenvalue associated to the fundamental Gaussian eigenvector, that is, to the case $(n_{1}=0,n_{2}=0)$, reads $\sqrt{\pi /2\sigma ^{2}}b_{0}^{+}b_{0}^{-}$.

We have therefore shown that one is able to master the number of supermodes
and the magnitude and sign of their eigenvalues by a proper choice of the
modulation amplitudes of the parametric multimode interaction kernel %
\eqref{Gaussian modulated}. In Figure \ref{cases12} we give
two examples of kernels, designed so as to lead to eigenvalues respectively proportional to $\left\{ 1,1,-1,-1\right\} $ and $\left\{ -4,1,1,1,1\right\} $. We note that the results shown in the figures have been obtained by numerically diagonalizing the
kernels, what gives additional support to our previous approximate
analytical treatment.

Our analysis has been restricted so far to the case of kernels with
symmetric Gaussians. However, it can be readily extended to the more physical
case of a kernel factorizing in  two Gaussian functions of variables $x+x^{\prime }$ and $x-x^{\prime }$ having unequal widths \cite{Patera2010,Wasilewski2006}. In
fact this is the most general situation as it is not always possible to
configure the OPO/OPA in such a symmetric way (in SPOPOs, for instance, for
typical situations one has a difference of one order of magnitude between
Gaussian widths). In this case the kernel we propose has again the
factorized form (\ref{FG}), now with 
\begin{equation}
K_{\pm }\left( x\right) =\mathrm{e}^{-\frac{1}{2}\sigma _{\pm
}^{2}x^{2}}\sum_{n=0}^{N_{\pm }}\,b_{n}^{\pm }\cos \left( \beta _{n}^{\pm
}x\right) ,  \label{modulation Gauss multimode}
\end{equation}%
instead of (\ref{Gaussian modulated}), where we allowed for different
widths, $\sigma _{\pm }^{-1}$, along the directions $x+x^{\prime }$ and $%
x-x^{\prime }$. In Appendix B we give the eigenvalues and eigenvectors of
this general case.

\section{Controlling the number of supermodes and the eigenvalues in actual
parametric devices}

The method we sketched out above can be practically applied to spatial or
temporal modes of the OPO/OPA since in both cases a factorized form (\ref{FG}%
) of the kernel is a very good approximation (it is even exact in the
``spatial case"), and both $K_{\pm }$ admit as well an approximated Gaussian
form, see e.g. \cite{Patera2010,Wasilewski2006,Bennink2002}. The point is
then how to implement in real devices the quasiperiodic modulations we
introduced in the previous Section.

We will divide our presentation in two Subsections. The first one deals with
what we call the continuous case, which includes both problems in which a
continuous boson representation is used (single-pass devices) or problems in
which, being that representation discrete, the kernel can be treated as a
continuous function (SPOPO \cite{Patera2010} and the self-imaging OPO \cite%
{Lopez2009}). The second Subsection deals with what we call the discrete
case, in which a discrete boson representation is used and a discrete
treatment of the problem is simpler, given the relatively small number of
modes involved, like an OPO tuned to a single transverse mode family \cite%
{Navarrete2009}.

\subsection{The continuous case}

\label{ContinuousCase}

As for the Gaussian form of the subkernels $K_{\pm }$ we note that both the
phase-matching function $D$ (\ref{sinc},\ref{Phi}) of the temporal case and
the diffraction function $\Delta $ (\ref{Delta}) of the spatial case can be
well approximated by the Gaussian $\mathrm{e}^{-\frac{1}{2}\tau
_{1}^{2}\left( \omega +\omega ^{\prime }\right) ^{2}-\frac{1}{2}\sigma
_{-}^{2}\left( \omega -\omega ^{\prime }\right) ^{2}}$. In the temporal case 
\cite{Patera2010,Wasilewski2006}
\begin{equation}
\tau _{1}=\frac{(k_{\mathrm{p}}^{\prime}-k_{\mathrm{s}}^{\prime}) l_{\mathrm{c}}}{2},
\label{tau1}
\end{equation}
$k_{\mathrm{p},\mathrm{s}}^{\prime }$ being the derivatives of the pump/signal wavenumber with
respect to frequency at phase matching, while in the spatial case $\tau
_{1}=0$. Hence if the pump spectral amplitude $\alpha _{\mathrm{p}}\left(
\omega \right) $ has a Gaussian shape $\mathrm{e}^{-\frac{1}{2}\tau _{%
\mathrm{p}}^{2}\omega ^{2}}$, $\tau _{\mathrm{p}}$ being the individual
pulse duration, the total kernel can be approximated by $K=\mathrm{e}^{-%
\frac{1}{2}\sigma _{+}^{2}\left( \omega +\omega ^{\prime }\right) ^{2}-\frac{%
1}{2}\sigma _{-}^{2}\left( \omega -\omega ^{\prime }\right) ^{2}}$ with $%
\sigma _{+}^{2}=\tau _{1}^{2}+\tau _{\mathrm{p}}^{2}$. If the pump has not a
Gaussian shape but can be expressed as a sum over $\cos $
functions, $\tau _{\mathrm{p}}=0$ (hence $\sigma _{+}^{2}=\tau _{1}^{2}$)
and the Gaussian kernel will be multiplied by those modulations (see below).

As the pump amplitude $%
\alpha _{\mathrm{p}}$ is a factor of $K_{+}$, see Eq. (\ref{freq coupling}), a simple and practical way to tailor the kernel is by shaping the pump, as we will see more precisely now. 

\subsubsection{Tailoring the pump}

We focus on the temporal case \cite{Valcarcel2006,Patera2010,Wasilewski2006}
and hence deal with pumps consisting of pulses. However the ideas put
forward below can be applied equally to the spatial case \cite{Lopez2009} by
substituting the pulse shapers we consider by amplitude masks or simply by
superposing different plane waves.

\paragraph{Using pulse shapers with harmonic spectral response:}

\begin{figure*}[t]
\includegraphics[width=\textwidth]{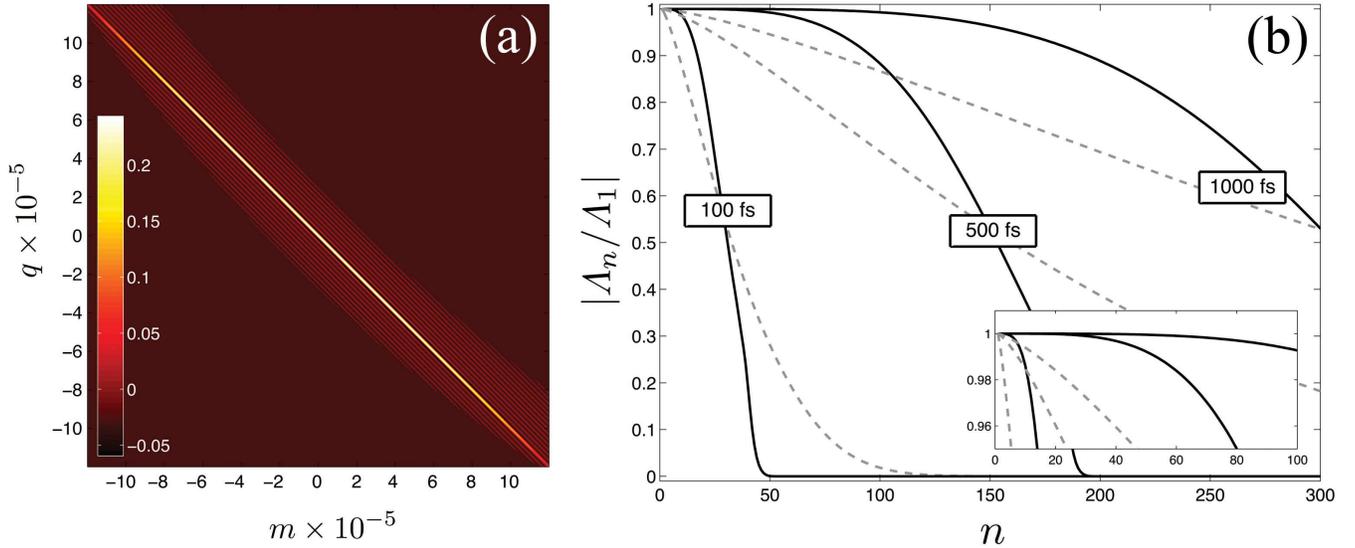}
\caption{(a) kernel evaluated in the realistic case of a $100\protect\mu$m
thick BIBO crystal inside a $4$m ring OPO cavity pumped by a c.w. train of $%
500$fs rectangular pump pulses for a degenerate type I critically phase
matching operation at $0.4\protect\mu$m pumping. $q$ and $m$ denote the
indices of the cavity longitudinal modes. (b) Eigenvalues obtained by
numerical diagonalization of the kernel shown in (a) when different trains
of pump pulses with Gaussian (grey-dashed lines) and rectangular (black-solid
lines) shapes are considered; the duration of the pulses are specified in the
figure. $\Lambda_1$ refers to the eigenvalue with the largest absolute value.
note that the caption is a zoom of the upper-left corner of the figure.}
\label{Ksinc}
\end{figure*}

A first possible method consists of using a pulse shaper \cite{Weiner2000}
with several well separated amplitude modulations on the Spatial Light
Modulator (SLM). The effect of such pulse shaper in the frequency domain can
be represented in the general case by the modulating function 
\begin{equation}
M(\omega )=\sum_{n=0}^{N}b_{n}\cos (\beta _{n}\omega ),
\end{equation}%
where $\beta _{0}=0$ by definition. The incoming pump field, of temporal
envelope $\alpha _{\mathrm{p,in}}(t)$, is sent onto a diffraction grating
and the image $\alpha _{\mathrm{p,in}}(\omega )$ is created in the Fourier
plane of a lens in correspondence of the SLM mask. Immediately after the
mask, the outgoing field has a spectrum $\alpha _{\mathrm{p,out}}(\omega
)=M(\omega )\alpha _{\mathrm{p,in}}(\omega )$. Then another lens and
diffraction gratings perform the inverse Fourier transform so that the
outgoing field has the following temporal shape: $\alpha _{\mathrm{p,out}%
}(t)=M(t)\otimes \alpha _{\mathrm{p,in}}(t)$, i.e. the usual convolution,
which is used to pump the SPOPO. It is then straightforward to see that, by
using $\alpha _{\mathrm{p,out}}(\omega )$ in Eq. \eqref{freq coupling}, one
obtains a modulation of the kernel $K(x,x^{\prime })$ as in Eq. 
\eqref{modulation Gauss multimode}, with $b_{n}^{+}=b_{n}$, $b_{0}^{-}=1$ and $b_{n>0}^{-}=0$. As a
consequence, the eigenvalues $\Lambda _{n}$ are proportional to $\pm b_{n}$
according to the analytical solutions given in Section \ref{Section4}.

\paragraph{Using pulse shapers leading to special waveforms:}

In this case we assume that a pulse shaper has been programmed so as to
convert the frequency comb coming from the pumping laser (whatever its
temporal waveform could be) into into a sequence of equal rectangular pulses at the frequency comb repetition rate. Hence the pump spectrum is proportional to $\mathrm{sinc}\left( \tau
_{\mathrm{p}}\omega \right) $, $\tau _{\mathrm{p}}$ being the pulse
duration. Then, if $\tau _{\mathrm{p}}^{-1}$ is much smaller than the width
of the crystal response function along the direction $\omega +\omega^{\prime }$---equal to $\tau _{1}^{-1}$ defined in Eq. \eqref{tau1}---, i.e. for
``long" pulses, the kernel can be approximated by $\mathrm{e}%
^{-\frac{1}{2}\sigma _{+}^{2}\left( \omega +\omega ^{\prime }\right) ^{2}-%
\frac{1}{2}\sigma _{-}^{2}\left( \omega -\omega ^{\prime }\right) ^{2}}%
\mathrm{sinc}\left[ \tau _{\mathrm{p}}\left( \omega +\omega ^{\prime
}\right) \right] $, see Eq. (\ref{freq coupling}), where $\sigma _{+}\sim
\tau _{1}$ \cite{Patera2010}. As a consequence, as $\mathrm{sinc}\left( \tau
_{\mathrm{p}}\omega \right) $ can be approximated by a finite (Fourier) sum
of functions $\cos \left( n\frac{2\pi }{L}\omega \right) $, with $L$ large
as compared to $\tau _{1}^{-1}$, the kernel reads as in (\ref{modulation Gauss multimode}). As the coefficients of the Fourier series, $b_{n}^{+}$, are all
very similar (because of the function $\mathrm{sinc}\left( \tau \omega
\right) $, whose Fourier transform is rectangular) we expect a large degree
of degeneracy between eigenmodes, as is actually evidenced by numerically
diagonalizing the corresponding kernel.

In Figure \ref{Ksinc}a we trace the kernel corresponding to a realistic
situation where a SOPO cavity, with the nonlinearity of a 100$\mu\mathrm{m}$-thick BIBO crystal ($\tau _{1}\sim 20$fs; see \cite{Patera2010} for
details), is pumped by a train of square pulses. The eigenvalues $\Lambda
_{n}$ obtained by the numerical diagonalization of $K$ for increasing values
of $\tau _{\mathrm{p}}$ are shown in Figure \ref{Ksinc}b and are compared
with the corresponding cases of Gaussian pump pulses: as expected, the degree
of degeneracy increases dramatically with the duration of the square pump
pulses. For instance, for $1\mathrm{ps}$ pulses the first hundred eigenvalues
differ by less than 1\%, see the inset in Figure \ref{Ksinc}b, what allows
the generation of very high quality, highly multi-dimensional entanglement.

\paragraph{Using delay lines}

A third method consists in using delay lines that, starting from one comb,
allow the superposition of several combs so that each \textquotedblleft
tooth" in the temporal domain is made of a series of pulses delayed/advanced
by $t_{n}$ with respect to the first one; one then gets as many eigenvalues
as there are different superposed combs, the eigenvalues being proportional
to the amplitude of each comb as we show next. For example, $\alpha (t)$
being the (normalized) envelope of the pump field associated to a train of
pulses incoming the OPA or the SPOPO cavity, the envelope of the pump field
associated to a generic superposition of delayed/advanced trains of pulses
reads 
\begin{equation}
\alpha _{\mathrm{p}}(t)=b_{0}\,\alpha (t)+\sum_{n=1}^{N}b_{n}\left[ \alpha
\left( t-t_{n}\right) +\alpha \left( t+t_{n}\right) \right] ,
\label{delayed pulses}
\end{equation}%
where $b_{0}$ and $b_{n}$ are real coefficients controllable in the
experiment. In the Fourier domain Eq. \eqref{delayed pulses} reads 
\begin{equation}
\alpha _{\mathrm{p}}(\omega )=\alpha (\omega )\left[ b_{0}+%
\sum_{n=1}^{N}b_{n}\cos (t_{n}\omega )\right] ,  \label{Ap delayed}
\end{equation}%
$\alpha (\omega )$ being the Fourier transform of $\alpha (t)$. Since we
have the same eigenproblem as in the previous cases, the eigenvalues $%
\Lambda _{n}$ will be again proportional to $\pm b_{n}$.

\subsubsection{Tailoring the crystal response}

Another interesting possibility in the temporal case is to modify not the pump temporal shape but
the effect of the nonlinear medium, which allows us to control the
function $K_{+}$. It can be accomplished by using several
identical non-linear crystals which are not perfectly phase-matched and separated by fixed distances. In the general
case where $N$ crystals of thickness $l_{\mathrm{c}}$ are used, centered at
planes $z=z_{n}$, the corresponding phase-matching function $D$, see (\ref%
{sinc}), reads%
\begin{eqnarray}
D(\omega ,\omega ^{\prime }) &=&\frac{1}{l_{\mathrm{c}}}\sum_{n}%
\int_{z_{n}-l_{\mathrm{c}}/2}^{z_{n}+l_{\mathrm{c}}/2}\mathrm{d}z\,\mathrm{e}%
^{\mathrm{i}\Delta k\left( \omega ,\omega ^{\prime }\right) z}
\\
&=&\sum_{n=1}^{N}\mathrm{\exp }\left[ \mathrm{i}\frac{2z_{n}}{l_{%
\mathrm{c}}}\Phi \left( \omega ,\omega ^{\prime }\right) \right] \mathrm{sinc%
}\left[ \Phi \left( \omega ,\omega ^{\prime }\right) \right] , \nonumber
\end{eqnarray}%
where $\Phi \left( \omega ,\omega ^{\prime }\right) $ is given in (\ref{Phi}%
). If the crystals are arranged symmetrically by couples at distances $d_{n}$
between mid-planes, then the above expression becomes%
\begin{equation}
D(\omega ,\omega ^{\prime })=2\sum_{n=1}^{N/2}\mathrm{\cos }\left[ 
\frac{d_{n}}{l_{\mathrm{c}}}\Phi \left( \omega ,\omega ^{\prime }\right) %
\right] \mathrm{sinc}\left[ \Phi \left( \omega ,\omega ^{\prime }\right) %
\right] ,  \label{DN}
\end{equation}%
where now the sum extends over couples of crystals\footnote{%
If the number of crystals is odd, say $N=2M+1$, then Eq. (\ref{DN}) is
modified by substituting the sum upper limit by $M$, and adding a term equal
to $\mathrm{sinc}\left[ \Phi \left( \omega ,\omega ^{\prime }\right) %
\right] $ to the result, as can be easily checked.}. When the pump spectrum
is narrow as compared with the width of $D$ along the direction $\omega+\omega ^{\prime }$---the quantity $\tau _{1}^{-1}$ defined in Eq. \eqref{tau1}---a safe
approximation consists in setting $\Phi \left( \omega ,\omega ^{\prime}\right) \rightarrow \tau _{1}\left( \omega +\omega ^{\prime }\right) $ in $\cos \left[ \frac{d_{n}}{l_{\mathrm{c}}}\Phi \left( \omega ,\omega ^{\prime}\right) \right] $. Hence,
\begin{equation}
\cos \left[ \frac{d_{n}}{l_{\mathrm{c}}}\Phi \left(\omega ,\omega ^{\prime }\right) \right] \approx \cos \left[ \frac{d_{n}}{l_{\mathrm{c}}}\tau _{1}\left( \omega +\omega ^{\prime }\right) \right],
\nonumber
\end{equation}
acts as a harmonic modulation (equivalent to $b_{n}^{+}=2$, $\beta _{n}^{+}=%
\frac{d_{n}}{l_{\mathrm{c}}}\tau _{1}$, $b_{0}^{-}=1$, and $b_{n>0}^{-}=0$
in our previous notation) of the single-crystal kernel, and the latter can
be approximated by a factorized Gaussian form as already discussed, so the
full kernel takes the form (\ref{modulation Gauss multimode}).

\subsection{Discrete case}

\label{DiscreteCase}

As commented in Section \ref{Intracavity}, the discrete representation is useful when dealing a low number of signal modes. A typical
configuration corresponding to this case is given by OPO with monochromatic pump,
whose cavity is tuned at the subharmonic to some transverse mode family
represented by the family index $f$ \cite{Navarrete2009}. Recall that family 
$f$ contains the $f+1$ Laguerre-Gauss modes $L_{(f-l)/2,\pm l}(\mathbf{r})$ with $l=l_{0},l_0+2,...,f$, being $l_{0}$ equal to $0$ for even families and $1$ for odd families.

In order to simplify the upcoming discussion, we assume that the the pump
beam has a coaxial cylindrical symmetry, in which case Orbital Angular
Momentum (OAM) conservation ensures that the down-converted photons must
have opposite OAMs. This implies that, if one uses the basis of
Laguerre-Gauss modes $\mathrm{TEM}_{pl}$, the parametric
down-conversion Hamiltonian takes the form \cite{Navarrete2009} 
\begin{equation}
\hat{H}_{\mathrm{I}}=\mathrm{i}\hbar g \sum_{l}\frac{\chi _{l}}{1+\delta _{0,l}%
}\hat{a}_{l}^{\dagger }\hat{a}_{-l}^{\dagger }+\mathrm{H.c.},
\end{equation}%
with%
\begin{equation}
\chi _{l}=2\pi \int_{0}^{+\infty }r\mathrm{d}r\alpha _{\mathrm{p}%
}(r)\left[ \mathcal{R}_{(f-l)/2}^{l}(r)\right] ^{2},  \label{OverLapp}
\end{equation}%
$\alpha _{\mathrm{p}}(r)$ and $\mathcal{R}_{(f-l)/2}^{l}(r)$ being the
transverse profiles of the (normalized) pump field and the Laguerre-Gauss modes 
\footnote{
Explicitly, we have 
\begin{equation}
\mathcal{R}_{p}^{l}(r)=\sqrt{\frac{2p!}{\pi (p+l)!}}\frac{1}{w}
\left( \frac{\sqrt{2}r}{w}\right)^{l}L_{p}^{l}
\left(\frac{2r^{2}}{w^{2}}\right) 
\mathrm{exp}\left( -\frac{r^{2}}{w^{2}}\right) ,
\end{equation}%
where $L_{p}^{l}(x)$ are the modified Laguerre polynomials and $w$ is the
spot size of the beam at the cavity waist.} 
at the cavity waist, respectively,
and $\hat{a}_{l}^{\dagger }$ the
creation operator associated to the Laguerre-Gauss mode $L_{(f-\left\vert
l\right\vert )/2,l}(\mathbf{r})$, which we abbreviate as $L_{l}(\mathbf{r})$
from now on because a fixed value of the family index $f$ is assumed as
explained.

The continuous boson operators $\hat{a}(\mathbf{r})$ defined on the
transverse plane can be expressed in terms of Laguerre-Gauss modal operators 
as%
\begin{equation}
\hat{a}(\mathbf{r})=\sum_{l}\frac{1}{1+\delta _{l,0}}\left[ L_{l}(\mathbf{r})%
\hat{a}_{l}+L_{-l}(\mathbf{r})\hat{a}_{-l}\right].
\end{equation}
The kernel $K(\mathbf{r},\mathbf{r}^{\prime })$ of Hamiltonian (\ref{GenHam}%
) has therefore the following expression:
\begin{equation}
K(\mathbf{r},\mathbf{r}^{\prime }) = \sum_{l} \frac{\chi _{l}}{1+\delta _{0,l}} \left[ L_{-l}(\mathbf{r}) L_{l}(\mathbf{r}^{\prime }) + L_{-l}(\mathbf{r}^{\prime }) L_{l}(\mathbf{r})\right] .  \label{K}
\end{equation}
This expression is close to the one assumed in (\ref{FG}). It can actually
be brought to that exact form by introducing the Hybrid Laguerre-Gauss modes
\begin{subequations}
\begin{eqnarray}
C_{l}(\mathbf{r})& =\frac{1}{\sqrt{2(1+\delta _{l,0})}}\left[ L_{l}(\mathbf{r%
})+L_{-l}(\mathbf{r})\right] ,
\\
S_{l}(\mathbf{r})& =\frac{1}{\mathrm{i}\sqrt{2(1+\delta _{l,0})}}\left[
L_{l}(\mathbf{r})-L_{-l}(\mathbf{r})\right] ,
\end{eqnarray}
\end{subequations}
which are respectively proportional to $\cos l\phi $ and $\sin l\phi $. The
kernel can then be written as%
\begin{equation}
K(\mathbf{r},\mathbf{r}^{\prime })=\sum_{H=C,S}\sum_{l}\chi _{l}H_{l}(%
\mathbf{r})H_{l}(\mathbf{r}^{\prime }).
\end{equation}%
In other words, we have just shown that the Hybrid Laguerre-Gauss modes $C_l$
and $S_l$ are the supermodes for the present OPO configuration. In addition,
both the \textquotedblleft sine\textquotedblright\ and \textquotedblleft
cosine\textquotedblright\ type modes have the same eigenvalue $\Lambda_{H,l}=\chi _{l}$ ($H=C,S$). Note that nothing prevents $\chi _{l}$ from being
complex, say $\chi _{l}=|\chi _{l}|\exp (\mathrm{i}\psi _{l})$, what means
that the supermodes with different $l$ can be squeezed along different
directions of phase space, direction $(\pi +\psi _{l})/2$ in particular for
modes $\{H_{l}(\mathbf{r})\}_{H=C,S}$. As for the number of available
supermodes $N$, it is given by the number of modes contained in family $f$,
that is, $N=f+1$.

\begin{figure*}[t]
\centering
\includegraphics[width=0.9\textwidth]{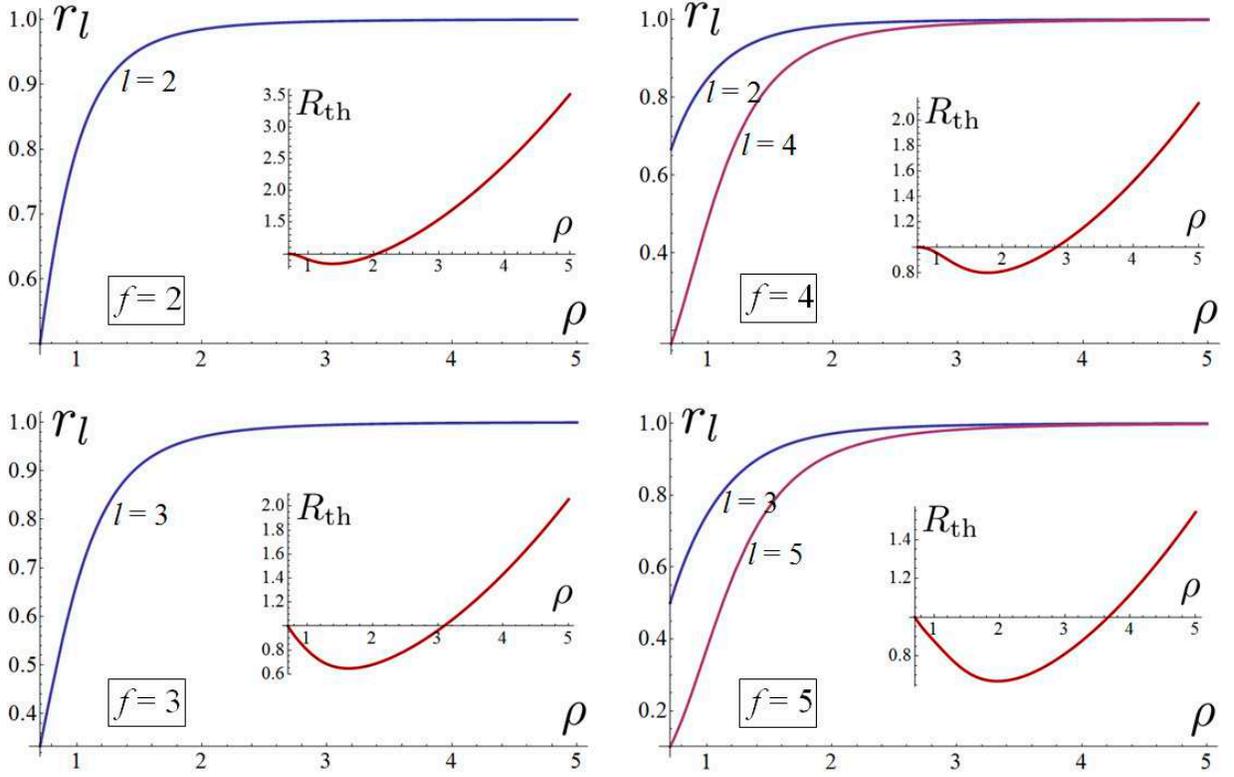}
\caption{Ratio between the couplings of the $l$ modes and the $l_{0}$ modes
as a function of the thickness of the Gaussian pump. The insets show the
corresponding change in the threshold pump power (normalized to the
threshold of the $\protect\rho=1/\protect\sqrt{2}$ situation), that is, the
power needed to make the signal field oscillate inside the cavity.}
\label{German1}
\end{figure*}

Let us now consider the situation in which the pump has a Gaussian shape,
that is, $\alpha_{\mathrm{p}}(r)= w_s G_{\mathrm{p}}(r)$, where 
\begin{equation}
G_{\mathrm{p}}(r)=\frac{1}{w_{\mathrm{p}}}\sqrt{\frac{2}{\pi }}\exp \left( -%
\frac{r^{2}}{w_{\mathrm{p}}^{2}}\right) ,
\end{equation}%
is a TEM$_{00}$ mode with spot size $w_{\mathrm{p}}=\rho w_{\mathrm{s}}$, $w_{%
\mathrm{s}}$ being the spot size of the transverse modes at the signal
frequency and $\rho $ a scaling factor (the factor $w_s$ in $\alpha_p$ is included for dimensional reasons, as we defined $\alpha_p$ as a dimensionless, normalized pump amplitude). It is obvious from (%
\ref{OverLapp}) that $\chi _{l_{0}}>\chi _{l_{0}+2}>...>\chi _{f}$; in other
words, the lower the OAM, the better the signal modes overlap with the pump
profile. However, for a large enough $\rho $, the pump profile is somehow
seen as plane by all the signal modes, and the couplings $\chi _{l}$ become
independent of $l$:%
\begin{equation}
\chi _{l}\underset{\rho \gg 1 }{\longrightarrow }2\pi \sqrt{\frac{2}{\pi}} \frac{w_s}{w_p} \int_{0}^{+\infty }rdr\left[ \mathcal{R}%
_{(f-l)/2}^{l}(r)\right] ^{2}=\sqrt{\frac{2}{\pi}} \rho^{-1}.
\end{equation}%
Hence, by pumping with a wide Gaussian, one can bring all the couplings $\chi _{l}$ to a common value, at the expense of decreasing them, hence increasing the OPO threshold. Note however that, as discussed next, the desired effect is obtained even for moderate values of $\rho$, so that the OPO threshold stays within reasonable limits.

In Fig. \ref{German1} we show the ratios $r_{l}=\chi _{l}/\chi _{l_{0}}$ as a function
of $\rho $ for the first families. Note that the $\rho $ axis starts at $1/%
\sqrt{2}$, which is the value one has in the doubly resonant configuration
(cavity resonant both for the signal and the pump). Note also how $r_{l}$
goes very quickly above $0.5$, which is the value above which one finds more
than 90\% squeezing for the corresponding supermodes \cite{Navarrete2009}.
In the insets, we show $R_{\mathrm{th}}(\rho )=\chi
_{l_{0}}^{2}(\rho =1/\sqrt{2})/\chi _{l_{0}}^{2}(\rho ),$ which gives the
ratio between the pump power needed to make the signal field oscillate for a
given $\rho $ and that for $\rho =1/\sqrt{2}$; note that the threshold is not
dramatically increased for reasonable values of $\rho $ (it is even
decreased for small $\rho $).

One can even tune the actual values of the couplings $\chi_l$ to specific values by pumping not with a single TEM$_{00}$ mode, but with a combination of TEM$_{00}$ modes of different widths.
To show this, let us consider two examples within the third family of transverse modes ($f=3$, and then $l=1,3$). First, we want to make $\chi_{1}=-\chi _{3}$, so that the eigenvalues of the $f+1=4$ supermodes have the same magnitude, but opposite sign. In this case, we can use the
following pump profile%
\begin{equation}
\alpha_{\mathrm{p}}(r)=w_s[G_{a}(r)\cos \theta -G_{b}(r)\sin
\theta ],
\end{equation}%
that is, a superposition of two Gaussians with opposite phase, and spot sizes 
$w_{a}=\rho _{a}w_{\mathrm{s}}$ and $w_{b}=\rho _{b}w_{\mathrm{s}}$. Note
that this type of pump shape can be easily obtained in the lab by mixing the
Gaussian beams on a beam splitter. It is then straightforward to show that choosing the mixing angle such that
\begin{equation}
\tan\theta=\left(
\frac{\rho_{a}}{\rho_{b}}\right)  ^{3}\left(  \frac{1+2\rho_{b}^{2}}%
{1+2\rho_{a}^{2}}\right)  ^{4}\frac{1+4\rho_{a}^{4}}{1+4\rho_{b}^{4}},
\end{equation}
one gets $\chi _{1}=-\chi _{3}$. Note that, in the general case where the resonator is tuned to family $f$, one can have complete control of all the ratios between the coupling parameters by using $1+ (f-l_{0})/2$ TEM$_{00}$ beams with adjustable amplitudes as pumping beams.

In the second example we show the very interesting case where some
of the couplings become zero by using an appropriate pump shape. To show that this extreme case is indeed
possible, we consider the previous example with the following choice of the
mixing angle between the Gaussians:%
\begin{equation}
\tan\theta=\left( \frac{\rho_{a}}{\rho_{b}}\right) ^{3}\left( \frac{%
1+2\rho_{b}^{2}}{1+2\rho_{a}^{2}}\right) ^{4}\frac{1+2\rho_{a}^{4}}{%
1+2\rho_{b}^{4}},
\end{equation}
in which case it is simple to show that $\chi_{1}=0$, while $\chi_3$ can be large enough with a proper election of  $\rho_{a,b}$. It can be interesting,
for example, to choose the mixing angle between the Gaussians in order to
cancel the coupling with the TEM$_{00}$ modes in a general OPO, so as to
favor the coupling to the TEM$_{10/01}$ modes, thus forcing it to emit the
signal and idler modes with opposite OAM. This way, one can induce a spontaneous breaking of the radial symmetry, what has been predicted to give rise to some remarkable quantum properties \cite{Navarrete08,Navarrete10}.

The methods discussed so far allow us to tune at will the eigenvalues of the
supermodes, but not the spatial profile of the supermodes themselves (they
are the Hybrid Laguerre-Gauss modes in all the cases). In order to change
the form of the supermodes, we can go a little further and add beams with
non-zero OAM to the pump, what would allow us to engineer any kind of
coupling between the modes with definite OAM of a given family. For example,
if one uses a pump field with a $+2$ OAM component, it becomes possible to
couple the signal modes with $+l$ and $-l+2$ OAM; mixing several of these
pump beams, one can even tune each of these couplings to a desired complex
value, so much as we have shown for the couplings between the signal modes
with $\pm l$ OAM by pumping with zero OAM Gaussian beams.

\section{Application to the generation of arbitrary cluster states}

\label{Cluster}

Having a set $\{s_{n}(\xi )\}_{n=1,2,...,N}$ of copropagating modes
with squeezing properties chosen at will (as we have discussed along the previous sections) offers the possibility of
generating any type of continuous-variable cluster state, that is, of
producing any type of Gaussian multipartite entangled state. To see this,
just note that we have proved that these modes, which we have called
supermodes, evolve according to the Hamiltonian%
\begin{equation}
\hat{H}_{\mathrm{I}}=\mathrm{i}\frac{\hbar g}{2}\sum_{n=1}^{N}\Lambda _{n}%
\hat{S}_{n}^{\dagger 2}+\mathrm{H.c.,}  \label{HPDC_diag}
\end{equation}%
where $\hat{S}_{n}^{\dagger }$ is the creation operator associated to the
supermode $s_{n}(\xi )$ of the kernel $K\left( \xi ,\xi ^{\prime }\right) $
having eigenvalue $\Lambda _{n}$, both of which can be controlled via any of
the ideas explained in the previous section. Note that this Hamiltonian can
be written as%
\begin{equation}
\hat{H}_{\mathrm{I}}=\mathrm{i}\frac{\hbar g}{2}\mathbf{\hat{S}}^{\dagger }%
\mathcal{L} (\mathbf{\hat{S}}^{\dagger }) ^{\mathrm{T}}+\mathrm{%
H.c.,}
\end{equation}%
where $\mathbf{\hat{S}}^{\dagger }=(\hat{S}_{1}^{\dagger },\hat{S}%
_{2}^{\dagger },...,\hat{S}_{N}^{\dagger })$ is a row vector operator,
superscript ``$\mathrm{T}$'' denotes transposition, and we have defined the diagonal matrix $\mathcal{L}=\mathrm{diag}\left( \Lambda _{1},\Lambda _{2},...,\Lambda _{N}\right) $. Through an arbitrary unitary matrix $\mathcal{U}$, we can define a
set of creation operators $\mathbf{\hat{B}}^{\dagger }=(\hat{B}_{1}^{\dagger
},\hat{B}_{2}^{\dagger },...,\hat{B}_{N}^{\dagger })=\mathbf{\hat{S}}%
^{\dagger }\mathcal{U}^{\dagger }$ for some new modes, so that 
\begin{equation}
\hat{H}_{\mathrm{I}} = \mathrm{i} \frac{\hbar g}{2}\mathbf{\hat{B}}^\dagger \mathcal{K} (\mathbf{\hat{B}}^\dagger)^\mathrm{T}+\mathrm{H.c.,}  \label{HPDC_G}
\end{equation}
where $\mathcal{K}=\mathcal{U} \mathcal{L} \mathcal{U}^{\mathrm{T}}$ is a new coupling matrix.
Hence (\ref{HPDC_diag}) can be seen as the diagonal form of (\ref{HPDC_G}).
For each choice of the coupling matrix $\mathcal{K}$, which can be chosen as
symmetric without loss of generality, this Hamiltonian generates a different
type of multipartite entanglement between the $\mathbf{\hat{B}}$ modes.

\begin{figure}[t]
\includegraphics[width=\columnwidth]{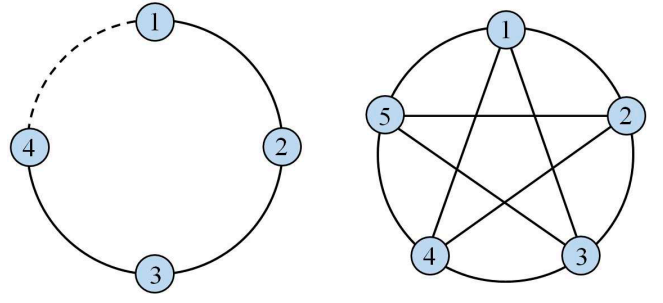}
\caption{Examples of cluster states. The dashed line denotes that the
corresponding coupling has opposite sign.}
\label{ClusterStates}
\end{figure}

As simple examples, consider the cluster states represented in Figure \ref%
{ClusterStates}. In the first type, four modes of a circle are connected
(all with same strength) only to their first-neighbors, one coupling having
opposite sign respect to the rest. The corresponding coupling matrix is (appart from a multiplicative factor)%
\begin{equation}
\mathcal{K}=\frac{1}{\sqrt{2}}%
\begin{bmatrix}
0 & 1 & 0 & -1 \\ 
1 & 0 & 1 & 0 \\ 
0 & 1 & 0 & 1 \\ 
-1 & 0 & 1 & 0%
\end{bmatrix}%
,
\end{equation}%
which has doubly degenerate
eigenvalues $\pm 1 $. Note that these eigenvalues correspond exactly to
some of the examples considered in previous sections, in particular,
to the example shown in Figure \ref{cases12}a of Section \ref{Section4}, and to
the second example of the previous section, where the couplings $\chi _{1}$
and $\chi _{3}$ of an OPO tuned tuned to the third family of transverse
modes were tuned to the same absolute value and opposite sign.

In the second example of Figure \ref{ClusterStates}, five modes are
interconnected with the same strength via the coupling matrix (again appart from a possible multiplicative factor)%
\begin{equation}
\mathcal{K}=-\frac{1}{4}%
\begin{bmatrix}
0 & 1 & 1 & 1 & 1 \\ 
1 & 0 & 1 & 1 & 1 \\ 
1 & 1 & 0 & 1 & 1 \\ 
1 & 1 & 1 & 0 & 1 \\ 
1 & 1 & 1 & 1 & 0%
\end{bmatrix}%
,
\end{equation}%
which has eigenvalues $\{-1 ,1 /4,1 /4,1 /4,1 /4\}$
coinciding with the example of Figure \ref{cases12}b in Section \ref{Section4}.
In the discrete spatial case, these eigenvalues can be obtained by tuning the
resonator to the $f=4$ family of transverse modes and tuning the couplings
to $\chi _{0}=-4\chi _{2}=-4\chi _{4}$, which can be done via the
multi-Gaussian pump technique discussed in the previous section, in this
case with $1+(f-l_0)/2=3$ Gaussians of appropriate widths and weights.

Incidentally, we note that these two cluster states are of very
different nature: while in the first case tracing out one of the modes does
not destroy the bipartite entanglement shared between the rest of the modes,
this is not the case in the second example, in which tracing out one of the
modes leads to a completely separable state. In the latter case, we then say
that the entanglement is ``genuinely multipartite'' \cite{Braunstein05}.

Indeed, the state of the second example in Figure \ref{ClusterStates} corresponds to the class of
so-called GHZ-like states \cite{Braunstein05}, that is, states which tend to%
\begin{equation}
|\mathrm{GHZ}\rangle _{N}=\int dx|\underset{N}{\underbrace{x,x,...,x}}%
\rangle ,
\end{equation}%
in the limit of perfect squeezing (kind of generalization of EPR states to $N$ modes), and are then completely characterized by the variance of the $\left( N+1\right)$ joint quadratures $\sum_{j=1}^{N}\hat{X}_{j}$ and $\left\{ \hat{P}_{j}-\hat{P}_{j+1\left( \mathrm{mod}N\right)}\right\} _{j=1\text{ }}^{N}$, being $\hat{X}_{j}=\hat{B}_{j}^{\dagger }+\hat{B}_{j}$ and $\hat{P}_{j}=\text{i}\left( \hat{B}_{j}^{\dagger }-\hat{B}_{j}\right) $ \cite{Braunstein05,vanLoock03}, of
which $|\mathrm{GHZ}\rangle _{N}$ is indeed an eigenstate.

\begin{figure}[t]
\centering
\includegraphics[width=\columnwidth]{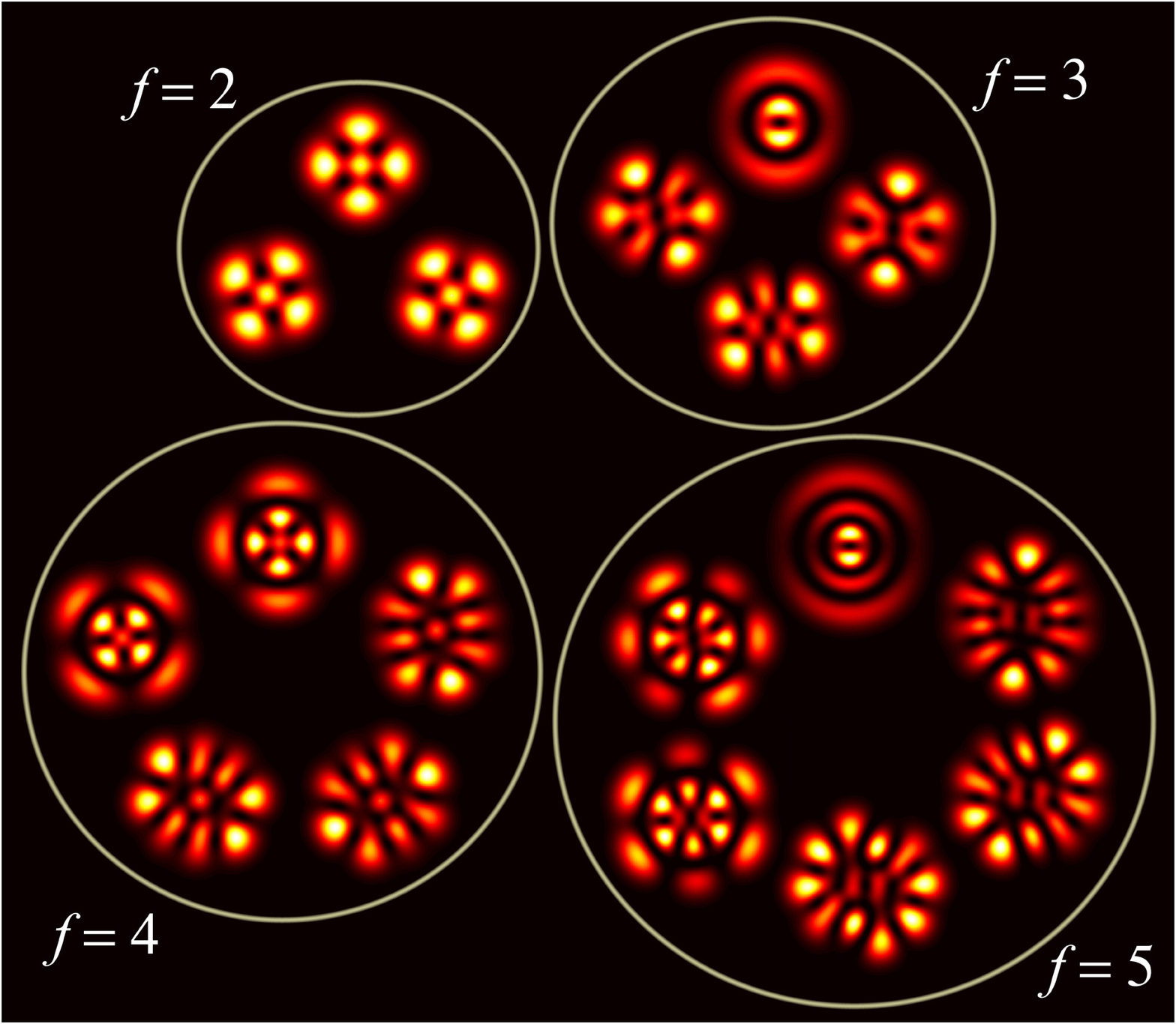}
\caption{Transverse profile of the modes sharing genuine multipartite
entanglement in the first families; they are obtained by applying a
Braunstein rotation onto the Hybrid Laguerre-Gauss modes (which act as the
supermodes in our system, that is, as independently squeezed modes).}
\label{MultiModes}
\end{figure}

There is another way of obtaining GHZ-like states from a set of $N$ squeezed
modes $\{\hat{S}_{n}\}_{n=1}^{N}$, which was introduced by van Loock and
Braunstein \cite{vanLoock00}, who proposed to start from $N-1$ modes
squeezed along a given direction of phase space, plus another one squeezed
along the orthogonal direction---a state that can be produced in the way
explained in the present paper---, and to apply to this state a specific
transformation known as the so-called ``Braunstein rotation" \cite%
{Braunstein1998,vanLoock00}.

Note that in order to have a reasonable approximation of the GHZ state, one
needs large squeezing in all the initial modes. In the previous section we
actually showed that by pumping with a wide enough Gaussian mode an OPO
tuned to the $f$-th family of transverse modes, one can bring together the
thresholds of the $f+1$ modes contained in that family, and hence all the
supermodes will leave the OPO highly squeezed when working close to threshold.
Note however that all the supermodes are squeezed in their $\hat{P}$ quadrature
(taking the pump phase as the reference), and hence, before applying
Braunstein's rotation, one needs to perform a $\pi /2$ phase shift in one of
them to obtain the true entangled modes. In Figure \ref{MultiModes} we
show the transverse profile (square modulus) of these modes for the first
families (the $\pi/2$ phase shift has been applied to the $C_{l_0}$ supermode). Note that using SPOPOs pumped by rectangular pulses we showed that
one can generate hundreds of highly squeezed supermodes (see Figure \ref%
{Ksinc}), and hence, this system offers a highly dimensional alternative to
the OPO tuned to a given transverse family where the number of squeezed
supermodes is more modest.

\section{Conclusions}

\label{Conclusions}

We have shown that an appropriate shaping of the pump beam (or of the geometry
of the nonlinear medium) enables us to generate in a multimode OPO any
Hamiltonian bilinear in the annihilation operators of the down-converted
modes, and therefore to generate highly multimode non-classical states of
light that may be of interest. In addition, we have shown that these quantum
states are easily characterized in a special mode basis, the basis of
``supermodes", as a superposition of independent, squeezed copropagating
modes.

The measurement of the quadrature components of any mode is always possible
with a balanced homodyne set-up using a coherent state in the mode of
interest as a local oscillator, which projects the multimode state on the
mode of the local oscillator. However, this technique is destructive, so
that it allows for a measurement of the properties of one supermode at a
time. An alternative to this single-mode homodyne detection is multiplexed
homodyne detection \cite{DelaubertThesis,Janousek09}, with which it is
possible to recover simultaneous information of the different supermodes
using an appropriate data processing protocol. This opens the
possibility of considering the copropagating entangled modes as a valid
resource for one-way quantum computation, at least in the Gaussian domain.

Even though having a cluster state between copropagating optical modes could
still be useful for certain quantum information processing tasks, a highly
relevant question which deserves further consideration is the way to access
the entangled modes independently (for example to perform measurements on
each of them, what is needed for one-way universal quantum computation), or
to separate them to further use in quantum communication networks.

There are different possible techniques to separate the different modes. For
example in the spatial domain, diffraction gratings are often used but
they introduce losses, which are detrimental for the quantum effects; combinations of spatial light modulators are promising candidates on this regard, as they are essentially unitary transforms on the
spatial modes \cite{Morizur2011}. In the time domain, the well-known pulse
shaping techniques can be of interest, but cannot transfer energy between
different frequency modes; on the other hand, light
modulators can do this task, and are the equivalent of diffraction gratings,
but they cannot modify the light spectrum by a large amount. Then, for temporal modes, the solution proposed in 
\cite{Brecht11} of using sum-frequency conversion with an appropriately
shaped pump beam is certainly the most promising.

\begin{acknowledgements}
We acknowledge fruitful discussions with E. Rold\'an and B. Chalopin. We
acknowledge the financial support of the Future and Emerging Technologies (FET)
programme within the Seventh Framework Programme for Research of the European
Commission, under the FET-Open grant agreement HIDEAS, number FP7-ICT-221906,
and of the Spanish Government and FEDER  through Projects FIS2008-06024-C03-01
and FIS2011-26960, and the FPU programme of the MICINN.
\end{acknowledgements}

\appendix

\section*{Appendix A}

The set of functions defined by (\ref{aux}) allows the diagonalization of
kernel (\ref{FG}) under (\ref{Gaussian modulated}). It is simple to show
that the action of the kernel on such functions, defined by the Fredholm
integral $F$ (\ref{Fredholm}), is%
\begin{equation}
F\left[ z_{m_{1},m_{2}}^{\mathrm{c},\mathrm{c}}\left( x\right) \right]
=\sum_{n_{1}=0}^{N_{+}}%
\sum_{n_{2}=0}^{N_{-}}b_{n_{1}}^{+}b_{n_{2}}^{-}Z_{m_{1},m_{2}}^{\mathrm{c},%
\mathrm{c}}\left( x\right) ,
\end{equation}%
where $\mathrm{c}=\cos $, $\mathrm{s}=\sin $,%
\begin{equation}
Z_{m_{1},m_{2}}^{\mathrm{c},\mathrm{c}}\left( x\right) =k_{\mathrm{c},%
\mathrm{c},m_{1},m_{2}}^{\mathrm{c},\mathrm{c},n_{1},n_{2}}z_{n_{1},n_{2}}^{%
\mathrm{c},\mathrm{c}}(x)+k_{\mathrm{c},\mathrm{c},m_{1},m_{2}}^{\mathrm{s},%
\mathrm{s},n_{1},n_{2}}z_{n_{1},n_{2}}^{\mathrm{s},\mathrm{s}}(x),
\end{equation}%
and%
\begin{eqnarray}
k_{t_{3},t_{4},m_{1},m_{2}}^{t_{1},t_{2},n_{1},n_{2}}=\int \mathrm{d}y\ 
\mathrm{e}^{-2\sigma ^{2}y^{2}}t_{1}\left( \beta _{n_{1}}^{+}y\right)
t_{2}\left( \beta _{n_{2}}^{-}y\right) \nonumber
\\
\times t_{3}\left( \beta
_{m_{1}}^{+}y\right) t_{4}\left( \beta _{m_{2}}^{-}y\right) ,
\label{kconst}
\end{eqnarray}%
are constants. When computing $F\left[ z_{m_{1},m_{2}}^{\mathrm{s},\mathrm{s}%
}(x)\right] $ a similar relation is obtained, showing that $\left\{
z_{m_{1},m_{2}}^{\mathrm{c},\mathrm{c}}\left( x\right) ,z_{m_{1},m_{2}}^{%
\mathrm{s},\mathrm{s}}\left( x\right) \right\} $ forms a closed set from
which actual eigenvectors can be found. Analogously the set $\left\{
z_{m_{1},m_{2}}^{\mathrm{c},\mathrm{s}}\left( x\right) ,z_{m_{1},m_{2}}^{%
\mathrm{s},\mathrm{c}}\left( x\right) \right\} $ is closed with respect to
the Fredholm integral (\ref{Fredholm}).

The method can be easily visualized in the simplest case where $b_{1}^{\pm }$
are the only non-null coefficients of the expansion. In such case, by
defining%
\begin{equation}
s\left( x\right) =f_{\mathrm{c}}z_{1,1}^{\mathrm{c},\mathrm{c}}\left(
x\right) +f_{\mathrm{s}}z_{1,1}^{\mathrm{s},\mathrm{s}}\left( x\right) ,
\end{equation}%
where $f_{\mathrm{c}}$ and $f_{\mathrm{s}}$ are constants to be determined, and computing $F\left[
s\left( x\right) \right] $ one obtains%
\begin{eqnarray}
F\left[ s\left( x\right) \right] =b_{1}^{+}b_{1}^{-} [\left( k_{\mathrm{%
c}}^{\mathrm{c}}f_{\mathrm{c}}+k_{\mathrm{s}}^{\mathrm{c}}f_{\mathrm{s}%
}\right) z_{1,1}^{\mathrm{c},\mathrm{c}}\left( x\right) \nonumber
\\
+\left( k_{\mathrm{c}%
}^{\mathrm{s}}f_{\mathrm{c}}+k_{\mathrm{s}}^{\mathrm{s}}f_{\mathrm{s}%
}\right) z_{1,1}^{\mathrm{s},\mathrm{s}}\left( x\right) ] ,
\end{eqnarray}
where $k_{\mathrm{s}}^{\mathrm{c}}=k_{\mathrm{s},\mathrm{s},1,1}^{\mathrm{c},\mathrm{c},1,1}$ and so on, are short notations for the constants introduced in (\ref{kconst}). Then the eigenvalue equation $F%
\left[ s\left( x\right) \right] =\Lambda s\left( x\right) $ is trivially
fulfilled by demanding%
\begin{equation}
\left. 
\begin{array}{c}
b_{1}^{+}b_{1}^{-}\left( k_{\mathrm{c}}^{\mathrm{c}}f_{\mathrm{c}}+k_{%
\mathrm{s}}^{\mathrm{c}}f_{\mathrm{s}}\right) =\Lambda f_{\mathrm{c}} \\ 
b_{1}^{+}b_{1}^{-}\left( k_{\mathrm{c}}^{\mathrm{s}}f_{\mathrm{c}}+k_{%
\mathrm{s}}^{\mathrm{s}}f_{\mathrm{s}}\right) =\Lambda f_{\mathrm{s}}%
\end{array}%
\right\} ,
\end{equation}%
which determines the eigenvalues $\Lambda $ by imposing that the system of
equations has nontrivial solutions. Two $\Lambda $'s are obtained ($\Lambda
_{1,2}$), proportional to the product $b_{1}^{+}b_{1}^{-}$, and for each of
them the ratio $f_{\mathrm{s}}/f_{\mathrm{c}}$ becomes fixed, which defines
the true eigenvectors, $s_{1,2}$ in this case.

\section*{Appendix B}

Although cumbersome it is straightforward to show that, if condition %
\eqref{approx} holds, the eigenvectors of the kernel (\ref{FG}) with Eq. %
\eqref{modulation Gauss multimode} are

\begin{equation}
s_{n_{1},n_{2},m}^{t_{1},t_{2}}(x)=\mathrm{e}^{-\tau ^{2}x^{2}}t_{1}\left(
\beta _{n_{1}}^{+}x\right) t_{2}\left( \beta _{n_{2}}^{-}x\right) \mathrm{H}%
_{m}\left( \sqrt{2}\tau x\right) ,
\end{equation}%
where $t_{i=1,2}$ stands for any of the trigonometric functions $\cos $ or $%
\sin $, $m\in \mathbb{N}$, $\tau =\sqrt{\sigma _{+}\sigma _{-}}$, and $%
\mathrm{H}_{m}\left( x\right) $ are Hermite polynomials. Their corresponding
eigenvalues read $\Lambda _{n_{1},n_{2},m}^{t_{1},t_{2}}=\lambda
_{m}b_{n_{1}}^{+}b_{n_{2}}^{-}$ when $t_{1}=\cos $ and $\Lambda
_{n_{1},n_{2},m}^{t_{1},t_{2}}=-\lambda _{m}b_{n_{1}}^{+}b_{n_{2}}^{-}$ when 
$t_{1}=\sin $, where 
\begin{equation}
\lambda _{m}=(-1)^{m}\frac{\sqrt{\pi /2}}{2\left( \sigma _{+}+\sigma
_{-}\right) }\left( \frac{\sigma _{+}-\sigma _{-}}{\sigma _{+}+\sigma _{-}}%
\right) ^{m}.
\end{equation}%
Hence also in this case one masters the number and eigenvalues of the
supermodes as well as their shapes. Note that for $\sigma _{+}=\sigma
_{-}=\sigma $ (symmetric Gaussians) $\lambda _{m\neq 0}=0$ while $\lambda
_{0}=\sqrt{\pi /32\sigma ^{2}}$, and one recovers the results of the
symmetric case. Note finally that when modulations are absent the previous
eigenvectors and eigenvalues coincide with those for a SPOPO pumped by
Gaussian pulses \cite{Patera2010}.

\end{document}